\documentclass[aps,prd,showpacs,amsmath,amssymb,reprint,superscriptaddress]{revtex4-1}
\usepackage{hyperref}
\usepackage{graphicx}
\usepackage{color}

\newcommand{\Par}{\mathcal{P}}
\newcommand{\fampl}{\mathcal{F}}
\newcommand{\gampl}{\mathcal{G}}

\begin{document}

\title{Inferring black hole charge from backscattered electromagnetic radiation}

\author{Lu\'is C. B. Crispino}
 \email{crispino@ufpa.br}
  \affiliation{Faculdade de F\'isica, Universidade Federal do Par\'a,
66075-110, Bel\'em, PA, Brazil}

\author{Sam R. Dolan}
\email{s.dolan@sheffield.ac.uk}
\affiliation{Consortium for Fundamental Physics,
School of Mathematics and Statistics,
University of Sheffield, Hicks Building, Hounsfield Road, Sheffield S3 7RH, United Kingdom.}

\author{Atsushi Higuchi}
 \email{atsushi.higuchi@york.ac.uk}
  \affiliation{Department of Mathematics, University of York,
Heslington,
York YO10 5DD, United Kingdom}

\author{Ednilton S. de Oliveira}
 \email{esdeoliveira@gmail.com}
 \affiliation{Faculdade de F\'isica, Universidade Federal do Par\'a,
66075-110, Bel\'em, PA, Brazil}

\date{\today}

\begin{abstract}
We compute the scattering cross section of
Reissner-Nordstr\"om black holes for the case of an
incident electromagnetic wave. We describe how scattering is affected by both the conversion
of electromagnetic to gravitational radiation, and the parity-dependence of phase shifts induced by the black hole charge. The latter effect creates a helicity-reversed scattering amplitude that is non-zero in the backward direction. We show that from the character of the electromagnetic wave scattered in the backward direction it is possible, in principle, to infer if a static black hole is charged.
\end{abstract}

\maketitle

\section{Introduction}
In 1909, the Geiger-Marsden (GM) experiment \cite{Geiger:Marsden:1909} revealed the internal structure of the atom, demonstrating the existence of a compact nucleus $\sim 10^4$ times smaller in diameter than the atom itself. Later, recalling the anomalous back-scattering of $\alpha$-particles from gold atoms, Rutherford remarked \cite{Rutherford:lecture} that ``it was almost as incredible as if you fired a 15-inch shell at a piece of tissue paper and it came back and hit you.''

Here we consider anomalous back-scattering in a rather different setting. In principle, one may infer the internal structure of astrophysical systems harboring compact nuclei -- such as black holes -- in a way analogous to the GM experiment [(i)] by observing the scattering of waves and particles on the curved spacetimes of General Relativity [(ii)]. Although a difference in scale prohibits laboratory-based experiments,  there are many formal similarities. For instance, long-ranged ($1/r$) effects dominate, but reveal little about internal structure: in case (i) a Coulomb field generates Rutherford scattering, which is insensitive to internal structure, and in case (ii) the `Newtonian' field component leads to an analogous scattering pattern, with an Einstein ring \cite{Einstein:1936} that is insensitive to the mass distribution. In either case, the `nucleus' attracts other matter/fields (i.e.~electrons, accretion disks, etc.) which may screen, in case (i), or distort, in case (ii), the scattering pattern. The `nucleus' itself generates weak effects on a narrower scale: for example, direct collisions in (i) and, in (ii), excitement of neutron star resonances, or in the black hole case, absorption and a host of effects associated with the `light ring' of radius $\sim 3 GM / c^2$, such as quasinormal ringing \cite{Chandrasekhar:Detweiler:1975}. In particular, the light-ring scatters flux through large angles, creating interference and a `glory' in the backward direction \cite{Matzner_etal}. Thus, back-scattered flux may provide telling hints about internal structure.

Several works have already been devoted to the topic of 
scattering by black holes (see Ref.~\cite{FHM-book} and references therein).
Many interesting black hole phenomena were 
explored in the 1970s, such as
superradiance~\cite{Starobinsky}, the glory
effect~\cite{prd10_1059}, and Hawking radiation \cite{Hawking}.  
It was realised that {\it charged} black holes provide an efficient mechanism for the conversion of electromagnetic (EM) to gravitational radiation and
vice versa~\cite{Zerilli,Gerlach,Moncrief,Olson-Unruh,Matzner}. 

Here we consider an EM wave incident upon a
Reissner-Nordstr\"om black hole. It was recently shown that the conversion mechanism plays an important role in absorption processes~\cite{cho, chm2010}, such that the gravitational and the EM absorption cross sections coincide in the extremal limit~\cite{cho2011}. 
Here we consider the scattering of EM flux. We show that charged black holes, unlike their Schwarzschild counterparts \cite{cdo2}, can scatter EM flux through exactly $180^\circ$, leading to a distinctive signature.

The Reissner-Nordstr\"om line element is given
by~\cite{Chandrasekhar}:
\begin{equation}
 ds^2=f(r)dt^2-f(r)^{-1}dr^2-r^2(d\theta^2+\sin^2\theta d\phi^2),
 \label{rn}
\end{equation}
where $f(r) = (1-r_{+}/r)(1-r_{-}/r)$, with
$r_{\pm} = M\pm\sqrt{M^2-Q^2}$,
and $M$ and $Q$ are the black hole mass
and charge, respectively.
We use natural units with $c = G = $1
and the metric signature $(+---)$ throughout.

It was shown in Ref.~\cite{cdo,Eiroa,Bhadra,Sereno} that, at small scattering angles, the scattering cross section is
\begin{equation}
\frac{d\sigma}{d\Omega} \approx \frac{16M^2}{\theta^4} +
\frac{3 \pi (5 M^2 - Q^2)}{4\theta^3} + O(\theta^{-2}) .
\label{weak_sca}
\end{equation}
Black hole charge leads to a sub-dominant correction to scattering in the weak-field region. Lensing properties, such as the Einstein ring, are dominated by the mass of the black hole, and the presence of charge is difficult to infer.

At larger scattering angles, the cross section will exhibit `orbiting' oscillations \cite{Anninos}, due to  interference between wavefronts which pass in opposite senses around the black hole. Such oscillations are signatures of the strong-field region of the spacetime. Near the backward direction ($\theta \sim 180^\circ$), such interference typically creates a `glory'. For massless fields of spin $s$, a WKB analysis gives \cite{Zhang, Matzner_etal}
\begin{equation}
\frac{d \sigma}{d \Omega} = 2 \pi \omega b_g^2 \left. \frac{d b}{d \theta} \right|_{b=b_g} J^2_{2s}( \omega b_g \sin \theta) .
\label{eq:glory}
\end{equation}
Here $b_g$ is the impact parameter associated with a null geodesic passing all the way around the black hole, near the light-ring. Note that the Bessel function $J_{2s}$ is zero at $180^\circ$ for massless fields with spin ($s>0$). Hence, according to Eq.~(\ref{eq:glory}) {\it no} flux will be scattered through exactly $180^\circ$. The semi-classical interpretation is that an annulus of the incident wavefront is focussed onto $\theta = 180^\circ$ and, since the spin is parallel-transported along each geodesic passing through the annulus, the circular symmetry results in complete destructive interference. 

However, even in uncharged ($Q=0$) black hole spacetimes, there is an additional feature in the back-scattering of {\it gravitational} waves ($s=2$) \cite{Chrzanowski:1976} that is not captured by Eq.~(\ref{eq:glory}). There arises a difference in the phase shift of the `odd' and `even' parity contributions (see below), which generates an additional scattering amplitude $\gampl(\theta)$ associated with the reversal of helicity \cite{FHM-book}. In the Schwarzschild case $|\gampl|^2 \sim M^2 \sin^{4}(\theta / 2)$ in the long-wavelength regime \cite{Matzner:Ryan:1977, Dolan:2008a}, implying a cross section of $M^2$ at $\theta = 180^\circ$. In the Kerr case, the backscattered flux may be greatly enhanced by superradiance, by a factor of up to $\sim 35$ times (cf.~Fig.~14 in Ref.~\cite{Dolan:2008b}). Below we show that a similar effect occurs for the scattering of purely EM waves by a Reissner-Nordstr\"om black hole. 

\section{Analysis}
Electromagnetic and gravitational perturbations in
Reissner-Nordstr\"om spacetime consist of axial (`odd-parity') and polar (`even-parity') modes~\cite{Olson-Unruh, Moncrief, Matzner}. The governing equations can be separated by parity $\Par = \pm$, with $+$ and $-$ denoting even and odd cases, respectively, leading to
decoupled ordinary differential equations \cite{Moncrief,Matzner},
\begin{equation}
 \frac{d^2}{dr_*^2}\varphi^{\Par}_{\pm} +
\left(\omega^2-V^{\Par}_{\pm}
 \right)\varphi^{\Par}_{\pm} = 0,
 \label{radial_eq}
\end{equation}
where $r_*$ is the Wheeler tortoise coordinate defined by
$
dr / dr_* = f 
\label{Tortoise}
$.
The $\pm$ signs appearing in subscripts in Eq.~(\ref{radial_eq}) are
related to the different expressions of the effective potentials
$V_{\pm}^{\Par}$, which are explicitly given in
Refs.~\cite{cho, cho2011}. We note that the modes $\varphi_+^\Par$ exist for $l \geq 1$,
and the modes $\varphi_-^\Par$ exist for $l\geq 2$.

Radial functions for the EM and gravitational waves, $F^{\Par}$ and $ G^{\Par}$ respectively, are given by
\begin{eqnarray}
 F^{\Par} &=& \varphi_{+}^{\Par}\cos \psi - \varphi_{-}^{\Par}
\sin \psi,
\label{F} \\
G^{\Par} &=& \varphi_{-}^{\Par}\cos \psi + \varphi_{+}^{\Par}
\sin \psi,
\label{G}
\end{eqnarray}
where
\begin{equation}
 \sin(2\psi) = -2 \Par Q\frac{\left[(l-1)(l+2) \right]^{1/2}}{\Omega},
 \qquad |\psi|<\frac{\pi}{4},
 \label{sin2psi}
\end{equation}
and
\begin{equation}
\Omega = \sqrt{ 9M^2 + 4Q^2(l-1)(l+2) } .
\end{equation}

The scattering of a pure EM wave corresponds to the asymptotic conditions (as $ r_* \to
\infty$)
\begin{eqnarray}
F(r_*) &\approx& F_{\omega l}^{\Par \text{in}} e^{-i\omega r_*} +
F_{\omega l}^{\Par \text{out}} e^{i\omega r_*} ,
\label{F_asy} \\
G(r_*) &\approx& G_{\omega l}^{\Par \text{out}} e^{i\omega r_*},
\label{G_asy}
\end{eqnarray}
where $F_{\omega l}^{\Par\text{in}}$, etc., are complex constants.
The radial functions
$\varphi_{\pm}^{\Par}$ have asymptotic forms
\begin{equation}
 \varphi_{\pm}^{\Par} (r_*) \sim \left\{
\begin{array}{lr}
e^{-i\omega r_*} + A_{\pm,\omega l}^{\Par}e^{i\omega r_*}, & (r_*
\to
\infty); \\
B_{\pm,\omega l}^{\Par} e^{-i\omega r_*}, & (r_* \to - \infty).
\end{array}
\right.
\label{phi_asy}
\end{equation}
which lead to 
\begin{eqnarray}
 R_{\omega l}^{\Par} &\equiv& \frac{F_{\omega
l}^{\Par \text{out}}}{F_{\omega l}^{\Par \text{in}}} =
A_{+,\omega l}^{\Par} \cos^2\psi + A_{-,\omega l}^{\Par}
\sin^2\psi ,
\label{Ref}  \\
 C_{\omega l}^{\Par} &\equiv& \frac{G_{\omega
l}^{\Par \text{out}}}{F_{\omega
l}^{\Par \text{in}}} = \frac{\sin(2\psi)}{2}(A_{+,\omega
l}^{\Par} - A_{-,\omega l}^{\Par}).
\label{Conv}
\end{eqnarray}
Here $|R_{\omega l}^{\Par}|^2$ and $|C_{\omega l}^{\Par}|^2$
represent the amount of reflected (non-converted)
and converted energy, respectively, when the incident
wave is purely EM. The polar and axial coefficients are related by \cite{Chandrasekhar}
\begin{equation}
\frac{A_{\pm,\omega l}^{+}}{A_{\pm, \omega l}^{-}} = \frac{(l-1)l(l+1)(l+2) + 2 i \omega \nu_{\mp}}{(l-1)l(l+1)(l+2) - 2i\omega \nu_{\mp}}
\label{odd-even-relation}
\end{equation}
where $\nu_{\pm} = 3M \pm \Omega$. 

The EM differential scattering cross section in spherically symmetric
spacetimes (for a circularly polarized incident planar wave) was found by Fabbri~\cite{Fabbri}
\begin{widetext}
\begin{equation}
 \frac{d\sigma}{d\Omega} = \frac{1}{8\omega^2} \left\{ \left|
\sum\limits_{l=1}^{\infty}
\frac{2l+1}{l(l+1)}\left[e^{2i\delta_{l}^{-}(\omega)}
T_l (\theta) + e^{2i\delta_{l}^{+} (\omega)}\pi_l (\theta) \right]
\right|^2 +
\left|\sum\limits_{l=1}^{\infty} \frac{2l+1}{l(l+1)} \left[
e^{2i\delta_{l}^{-} (\omega)} \pi_l (\theta) +
e^{2i\delta_{l}^{+} (\omega)} T_l (\theta) \right] \right |^2
\right\},
\label{scs}
\end{equation}
\end{widetext}
with the phase shifts given by 
\begin{equation}
e^{2 i \delta_l^{\Par}(\omega)} = (-1)^{l+1} R_{\omega l}^{\Par} ,
\label{phase_shift}
\end{equation}
and angular functions given by
\begin{equation}
 \pi_l(\theta) \equiv \frac{P_l^1(\cos\theta)}{\sin\theta}, \qquad
T_l (\theta) \equiv \frac{d}{d\theta } P_l^1 (\cos\theta),  \label{eq:pi-T}
\end{equation}
where $P_l^m (\cos\theta)$ are associated Legendre functions.
Eq.~(\ref{scs}) may be recast as $\frac{1}{2} \left(|\fampl + \gampl|^2 + |\fampl - \gampl|^2 \right) = |\fampl|^2 + |\gampl|^2$, with
\begin{eqnarray}
\fampl(\theta)  &=& \frac{\pi}{i \omega} \sum_{l=1}^\infty \sum_{\Par = \pm} \left[ \exp \left( 2 i \delta_{l}^\Par \right) - 1  \right] {}_{-1}Y_l^1(1) {}_{-1}Y_l^1(\cos\theta) ,  \nonumber \\
\gampl(\theta) &=& \frac{\pi}{i \omega} \sum_{l=1}^\infty \sum_{\Par = \pm} \left[ \exp \left( 2 i \delta_{l}^\Par \right) - 1  \right] \Par (-1)^l \times \nonumber \\ 
&& \quad \quad \quad \quad \quad \quad {}_{-1}Y_l^1(1) {}_{-1}Y_l^1(- \cos\theta) .  \label{eq:fg}
\end{eqnarray}
Here ${}_sY_l^m(\cdot)$ are the spin-weighted spherical harmonics \cite{Goldberg}, and $\fampl$ and $\gampl$ are the helicity-preserving and helicity-reversing amplitudes. A consequence of Eq.~(\ref{odd-even-relation}) is that $\delta_l^{+} \neq \delta_{l}^{-}$, except for in the Schwarzschild case; hence $\mathcal{G} \neq 0$ for charged black holes. We note that $\fampl(\theta = 180^\circ) = 0$ by construction, whereas $\gampl(\theta = 180^\circ) \neq 0$.  

By applying the method of Ref.~\cite{Dolan:2008a} we may obtain approximations for $\fampl$ and $\gampl$ in the low-frequency regime. We find
\begin{equation}
|\gampl|^2 = q^4 M^2 \sin^4(\theta / 2) + O(\omega^2) \label{eq:lowfreq}
\end{equation}
with $q \equiv |Q|/M$. At low frequencies the only contribution to $\gampl$ is from the $l=1$ mode. This shows that helicity-reversal and electromagnetic-to-gravitational conversion are distinct phenomena, as the latter occurs only for $l \ge 2$.

To compute the cross section for general frequencies, we used numerical methods. First, we computed the phase shifts by solving the radial
equations numerically and matching the solutions onto their analytical
asymptotic forms. Second, we used a convergence method to sum the (formally divergent) partial-wave series (\ref{scs}).

For the first step, the asymptotic forms~(\ref{phi_asy}) are not sufficiently accurate. Instead, we write the asymptotic form of the solutions to the radial equations~(\ref{radial_eq}) in terms of spherical Hankel functions $h_l^{(1)}$, which are obtained by keeping only the term $l(l+1)/r^2$ in the effective potentials of Eq.~(\ref{radial_eq}).  For the second step, we adapted the method first introduced by Yennie \textit{et al.}~\cite{Yennie}, which has been successfully applied to e.g.~the study of scalar scattering by Reissner-Nordstr\"om black holes~\cite{cdo}.

\begin{figure*}
\centering
\includegraphics[width=8.6cm]{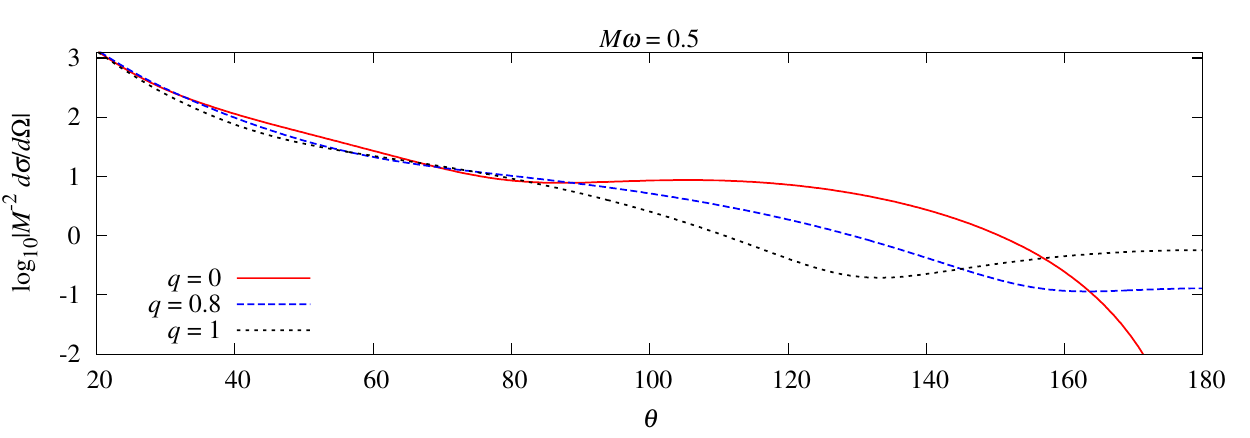}
\includegraphics[width=8.6cm]{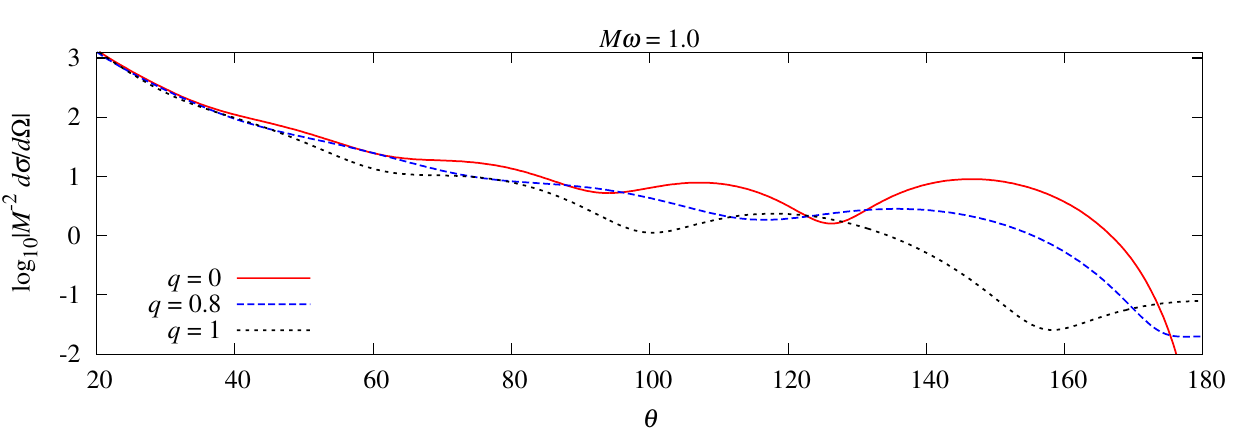}
\includegraphics[width=8.6cm]{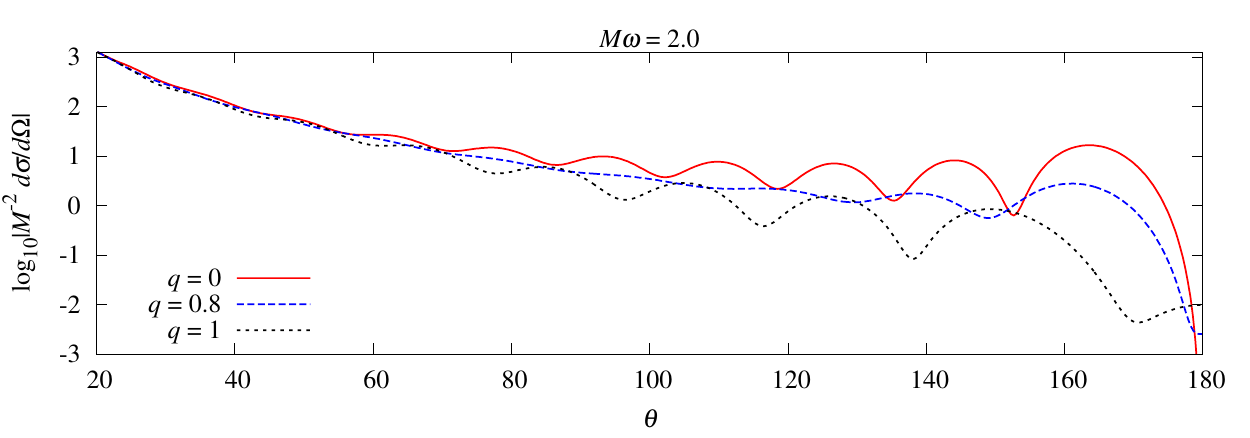}
\includegraphics[width=8.6cm]{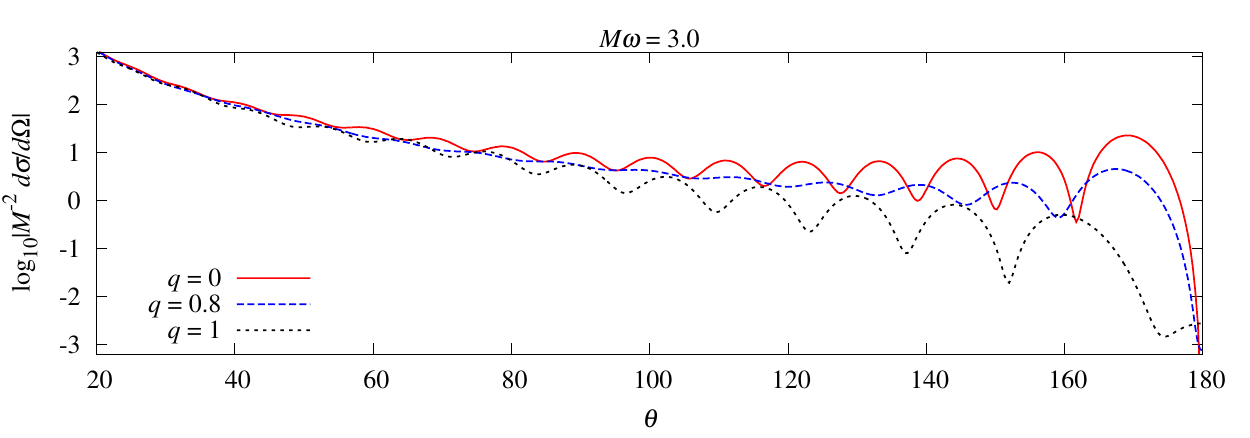}
\caption{Electromagnetic scattering by Reissner-Nordstr\"om black
holes for $q = 0, 0.8, 1$ and $M\omega = 0.5, 1.0, 2.0, 3.0$. For $0 < q \le 1$, the flux of EM radiation in the backward direction is non-zero; it diminishes as $M \omega$ increases.
}
\label{rn-sca}
\end{figure*}

Figure~\ref{rn-sca} shows the scattering cross sections obtained
numerically for the cases $q = 0, 0.8, 1$ and $M\omega = 0.5, 1, 2, 3$.
Increasing the charge-to-mass ratio $q$ at fixed $M\omega$ leads to wider `orbiting' oscillations, and a smaller (average) flux at large angles. 

\begin{figure}
 \centering
 \includegraphics[width=8.6cm]{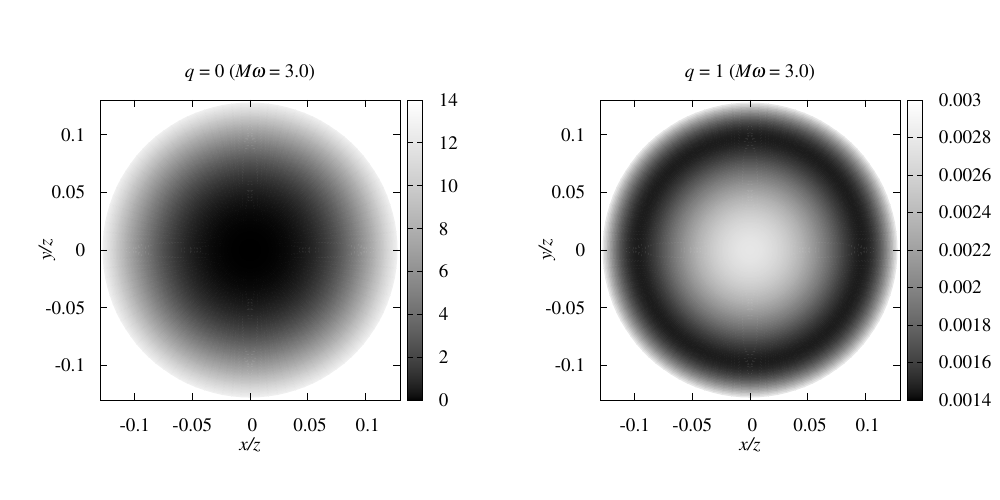} 
 \caption{Illustration of an electromagnetic-wave detection in the backward direction for a Schwarzschild ({\it left}) and a extreme Reissner-Nordstr\"om ({\it right}) black hole. Here, $M\omega = 3.0$  and the scattering angle interval is
 between $172.6^\circ$ and $180^\circ$ in both graphs.
 }
 \label{fig:brightspot}
\end{figure}

A novel feature of scattering of EM waves by
Reissner-Nordstr\"om black holes is the appearance of flux on-axis in the backward direction for $0 < q \le 1$. This effect, although small, can be seen clearly in Fig.~\ref{rn-sca}. Figure~\ref{fig:brightspot} illustrates the angular profile of this flux, as it might be seen in a detector. A bright spot of flux around $180^\circ$ in the extreme case (right) contrasts with the dark spot in the Schwarzschild case (left). 

\section{Discussion}
We have explored a novel signature of black hole charge: the non-zero flux around $180^\circ$ when a black hole scatters EM plane waves. We have shown that this effect arises from parity-dependence in the scattering phase shifts, leading to a `helicity-reversed' scattering amplitude $\gampl(\theta)$. At low frequencies, the helicity-reversed flux is primarily in the $l=1$ mode, and the cross section at $180^\circ$ is $\sim Q^4 / M^2$ [cf.~Eq.~(\ref{eq:lowfreq})]. Note that the back-scattered flux is dissimilar to a backward `glory', as it diminishes as $M\omega$ is increased and is primarily a low-frequency, low-multipole effect.

Parity-dependence in scattering is related to the presence of a source for the field. On one hand, the black hole's mass acts as a source for the gravitational field, leading to parity dependence in the gravitational sector. Parity-dependence for gravitational waves on uncharged spacetimes was noted many years ago, and was shown to lead to non-zero backscattering in Refs.~\cite{Westervelt,Peters,DeLogi-Kovacs}. On the other hand, the black hole's charge acts as a source for the EM field, generating parity-dependence and backscattering in the EM sector. In the low-frequency limit the helicity-reversed cross sections take a similar form, with $|\gampl|^2 \sim M^2 \sin^4 (\theta / 2)$ and $|\gampl|^2 \sim q^4 M^2 \sin^4 (\theta / 2)$ in the gravitational and EM cases, respectively; in the extremal limit $q \rightarrow 1$ these become identical.

We note that helicity-reversal and electromagnetic-to-gravitational conversion are two rather distinct effects. The former arises from parity dependence as exhibited by Eq.~(\ref{odd-even-relation}). The latter arises from mixing between EM and gravitational sectors, via Eq.~\eqref{sin2psi}. The latter, conversion, will generate other interesting effects in scattering, such as the generation of EM flux from gravitational-wave irradiation, and {\it vice versa}.

In principle, as described here, observations of backscattered EM flux allow one to infer the black hole's charge. In practice, the backscattering effect is likely to be negligible in astrophysical black hole scenarios due to its low-frequency character. In this regard it is somewhat like the Hawking effect. On the other hand, it is known that, for a rapidly-spinning black hole, the backscattering of gravitational waves is greatly enhanced by superradiance~\cite{Dolan:2008b}. We anticipate that EM back-scattering will be similarly enhanced in the Kerr-Newman black hole case. Thus, we conclude that the back-scattering effect is an interesting aspect of black hole phenomenology that warrants further investigation.

\acknowledgments
The authors are grateful to Conselho Nacional de Desenvolvimento
Cient\'\i fico e Tecnol\'ogico (CNPq) and to Coordena\c{c}\~ao de
Aperfei\c{c}oamento de Pessoal de N\'\i vel Superior (CAPES)
 for partial financial support.
A.~H.\ and L.~C.\ also acknowledge partial support from the
Abdus Salam International Centre for Theoretical Physics
through the Visiting Scholar/Consultant Programme and
Associates Scheme, respectively.
A.~H.\ and S.~D.\ thank the Universidade Federal do Par\'a (UFPA)
in Bel\'em for kind hospitality.

\end{document}